\documentclass[preprint,nofootinbib,APS]{revtex4}
\usepackage{amsmath,amssymb}
\usepackage{graphicx}
\usepackage{bm}

\begin{document}

\title{
Dynamical Compactification and Inflation \\ 
in Einstein-Yang-Mills Theory 
with Higher Derivative Coupling
}
\author{Hironobu Kihara$^{\dag}$}
\author{Muneto Nitta${}^{\ddag}$ }
\author{Misao Sasaki$^{\dag,\#}$}
\author{Chul-Moon Yoo${}^{*}$ }
\author{Ignacio Zaballa$^{\dag}$}

\preprint{APCTP-Pre209-003}
\preprint{KIAS-P09016}
\preprint{YITP-09-32}

\affiliation{${}^{\dag}$Korea Institute for Advanced Study\\
207-43 Cheongnyangni 2-dong, Dongdaemun-gu, 
Seoul 130-722, Republic of Korea
}
\affiliation{${}^{\ddag}$Department of Physics, 
Keio University, Hiyoshi, Yokohama, Kanagawa
 223-8521, Japan}
\affiliation{${}^{\#}$Yukawa Institute for Theoretical Physics\\
Kitashirakawa Oiwake-Cho, 606-8502 Kyoto, Japan
}
\affiliation{ ${}^{*}$ Asia Pacific Center for Theoretical Physics\\
Hogil Kim Memorial Building \# 501, POSTECH, San 31, 
Hyoja-dong, Namgu, Pohang, Gyeongbuk 790-784, Republic of Korea }

\date{\today}

\begin{abstract}
We study cosmology of the Einstein-Yang-Mills theory in ten dimensions
with a quartic term in the Yang-Mills field strength. 
We obtain analytically a class of cosmological solutions
in which the extra dimensions are static
and the scale factor of the four-dimensional 
Friedmann-Lemaitre-Robertson-Walker metric 
is an exponential function of time. 
This means that the model can explain inflation.
Then we look for solutions that describe dynamical compactification
of the extra dimensions. 
The effective cosmological constant $\lambda_1$ in the four-dimensional
universe is determined from the gravitational coupling, ten-dimensional
 cosmological constant, gauge coupling and higher derivative coupling. 
By numerical integration, the solution  with $\lambda_1=0$ 
is found to behave as a matter-dominated universe which asymptotically
approaches flat space-time, while the solution with a non-vanishing
 $\lambda_1$ approaches de Sitter space-time in the asymptotic future.
\end{abstract}

\maketitle

\tableofcontents
\section{Introduction}


There have been many attempts to consider extra dimensions 
in addition to our world of four-dimensional space-time, 
even though they have not been observed. 
The original idea dates back to Nordstr\"om~\cite{Nordstrom:1988fi},
Weyl \cite{Weyl:1918ib},
Kaluza~\cite{Kaluza:1921tu} and Klein~\cite{Klein:1926tv},
who considered extra dimensions in order to 
unify gravity and electromagnetic force in five space-time dimensions.
Now the most promising unified theory, 
describing all fundamental forces including 
two types of nuclear forces, 
is considered to exist in ten, eleven or twelve dimensions
 after string theory, M-theory ~\cite{Witten:1995ex}
 and F-theory ~\cite{Vafa:1996xn}
 have appeared.
Superstring theory is consistent in ten-dimensional space-time. 
The extra six dimensions should be compactified. 
Some people require supersymmetry in four-dimensional space-time and
 the extra-dimensional space was assumed as a Calabi-Yau manifold. 
After the discovery of D-branes, D-branes or more generally
``branes" offer the possibility of large extra dimensions or 
the brane-world scenario \cite{ArkaniHamed:1998rs}. 

There were many efforts to describe cosmological solutions 
in the framework of higher-dimensional theories. 
Especially, the realization of de-Sitter like expansion of 
a 4-dimensional part has attracted much attention 
in connection with the inflationary scenario or 
the current accelerated expansion of our universe. 
One of those attempts is the flux compactification, 
which have received a lot of attention in recent years~\cite{KKLMMT}. 
One of the most important and basic features of the flux compactification 
is to stabilize the size of a compactified space 
by certain configurations of high-rank differential form fields. 

Before string theory was discovered, 
Cremmer and Scherk studied an attractive possibility of 
compactification with the size of a compactified space 
being stabilized~\cite{Cremmer:1976ir}.  
In order to achieve it they placed a non-trivial topological 
solution (soliton) of a gauge field on the compactified space, 
for instance a monopole on the sphere $S^2$  
or a Yang-Mills instanton on the four-dimensional sphere $S^4$.  
In these cases, the compactified space is stabilized
at a finite radius rather than decompactified to an infinite radius.
So they called it ``spontaneous compactification".

In this paper we would like to study if 
such a compactification can occur dynamically or not. 
 In general, in order to stabilize a topological configuration
 of a Yang-Mills field in dimensions greater than four, 
we need higher order terms of the gauge field strength \cite{Derrick:1964ww}. 
Some years ago Tchrakian introduced such a term, 
which we call the Tchrakian term, 
in order to generalize 't Hooft-Polyakov monopoles and Yang-Mills instantons 
to those analogues in dimensions greater than four \cite{Tchrakian:1978sf}. 
The term is not renormalizable, but still quadratic in the time derivative.
Recently some of the present authors have numerically 
studied a monopole-like solution in six-dimensional Minkowski space  
by adding the Tchrakian 
term~\cite{Kihara:2004yz}.\footnote{This was originally motivated 
by the computation of non-Abelian Berry's phases 
in T-dualized USp matrix model \cite{Chen:1998qb}.}
One of the authors has further studied 
asymptotic solution of five-dimensional Tchrakian monopole, 
the generalization of Tchrakian monopole \cite{Kihara:2008fz}. 
In the case of a six-dimensional sphere, an exact solution to
a generalized self-duality relation has been constructed 
for SO(6) Yang-Mills fields 
with the Tchrakian term~\cite{Kihara:2007di}.\footnote{
 Generalization of instantons on the complex projective 
space ${\mathbb C}{\bf P}^3$  has been also given \cite{Kihara:2008zg}.
}
Then this relation has been successfully embedded in the 
Einstein-Yang-Mills theory with the Tchrakian term
 in the geometry of the direct product of the four-dimensional Minkowski space 
(anti-de Sitter space AdS$_4$) and $S^6$ of a constant radius,
with (without) a ten-dimensional cosmological constant~\cite{Kihara:2007vz}. 
In this solution the gauge field distributes on $S^6$ homogeneously, 
so it is a natural generalization of Cremmer and Scherk~\cite{Cremmer:1976ir}.
At least for the Yang-Mills part, the configuration 
attains the minimum of the Bogomol'nyi bound 
when the radius of $S^6$ satisfies a certain relation with  
the gauge coupling constant and the coupling strength of the Tchrakian term.
Therefore we expect that if we turn on the time variation
of the space-time, we obtain a solution which describes the
process of dynamical compactification.

In this paper 
we consider cosmological solutions with a time-dependent scale
factor of the three dimensions as well as with a time-dependent
radius of $S^6$, and study if there exist
solutions with the radius of $S^6$ tending to a finite value,
as a possible model of dynamical compactification.

This paper is organized as follows. 
 In Sec.~\ref{section:set}, we describe our theory, that is, 
the Einstein-Yang-Mills theory with the Tchrakian term in ten dimensions. 
We review the discussion of Bogomol'nyi completion.
In Sec.~\ref{section:FCSI}, we introduce an ansatz on the ten-dimensional
metric, namely, the direct product of the four-dimensional 
Friedmann-Lemaitre-Robertson-Walker (FLRW) metric and $S^6$ with the radius
as a function of time. Then we specify a gauge configuration
which satisfies the self-duality relation and 
solves the Yang-Mills equation with the Tchrakian term.
In Sec.~\ref{section:simple}, simple analytical solutions
with a fixed radius of $S^6$ are given. The four-dimensional part
of the solutions is either Minkowski or de Sitter, depending on
 the choice of the model parameters. 
In Sec.~\ref{section:dynamical}, we consider solutions that describe
the process of dynamical compactification. We investigate the behavior
of the solutions both analytically and numerically. In general,
the four-dimensional part behaves as de Sitter plus small oscillations,
while the $S^6$ radius undergoes damped oscillations toward a finite value.
For a particular choice of the model parameters that gives the product
of a flat space-time times $S^6$ with a fixed radius,
we find the four-dimensional part behaves 
as a dust-dominated universe, that is, with the scale factor proportional
to $t^{2/3}$.
Sec.~\ref{section:conclusion} is devoted to conclusion and discussions. 


\section{Model Setting and Bogomol'nyi Equation}
\label{section:set}

Let us start from the following action in ten-dimensional space-time:
\begin{align}
S_{{\rm tot}} &:= S_{_{\rm EH}} + S_{_{\rm YMT}} ~~,\cr
S_{_{\rm EH}}  &:=  \frac{1}{16 \pi G} \int  dv {\mathcal R} ~~, \cr
S_{_{\rm YMT}} &:= \frac{1}{16} \int {\rm Tr} 
\left\{ -  F \wedge * F +  \alpha^2 (F \wedge F) \wedge * (F \wedge F)
 - V_0 dv \right\}  ~~ .
\end{align}
Here $dv$ is the invariant volume form, ${\mathcal R}$ is the scalar curvature
 with respect to the metric $g_{MN}$ and $F$ is the field strength two-form
 which takes values in the Lie algebra 
so(6).   
The star ($*$) denotes the Hodge dual  operator acting on differential forms
in ten dimensions.
 Our notation is summarized in Appendix~\ref{sec:notation}. For more details, 
see \cite{Eguchi:1980jx}.

We consider the case where the space-time is locally
 a product space of ${\cal M}$ and ${\cal N}$. ${\cal M}$ is 
a four-dimensional curved space-time and ${\cal N}$ is a compact space.
Let us denote  the total space ${\cal T}$. 
Metric on this space is 
\begin{align}
ds^2 &= g_{\mu\nu}(x) dx^{\mu}dx^{\nu}
 + g_{IJ}(x,y) dy^I dy^J = ds^2_{{\cal M}} + ds^2_{{\cal N}}~,\cr
\mu, \nu &= 0,1,2,3~,~~
I,J = 4,5, \cdots , 9~.
\end{align}

For the case where the field strength has only components along the 
compact directions, we can manipulate the Yang-Mills action
as \cite{Kihara:2007vz}
\begin{align}
& \frac{1}{16} \int_{\cal T} {\rm Tr} \left\{ -  F \wedge * F
 +  \alpha^2 (F \wedge F) \wedge * (F \wedge F) \right\}  \cr
&\quad=  \frac{1}{16} \int_{\cal M} dv^{(4)} \int_{\cal N} {\rm Tr} 
\left[ \Bigl(F \mp {\bm{i}} \alpha \gamma_7 *_6 (F \wedge F) \Bigr) 
\wedge *_6  \Bigl(F \mp {\bm{i}} \alpha \gamma_7 *_6 (F \wedge F)\Bigr) 
 \right]\cr
&\qquad
\pm \frac{1}{16} \int_{\cal M}  dv^{(4)} \int_{\cal N}
 {\rm Tr}\,2 {\bm{i}} \alpha \gamma_7 F \wedge F \wedge F   ~,
\label{eqn:bogom}
\end{align}
where $*_6$ represents the Hodge dual operator along 
the compact direction ${\cal N}$. 
We call this procedure Bogomol'nyi completion. 
The term $Q := \int_{\cal N} {\rm Tr} \gamma_7 F^3$ is a surface term 
and it gives the bound on the energy density. 
Then the Bogomol'nyi equation is 
\begin{align}
F \mp {\bm{i}} \alpha \gamma_7 *_6 (F \wedge F) =0 ~.
\label{eqn:bogomolnyi}
\end{align}
If either of these equations is satisfied, the energy attains 
the minimum given by $Q$ irrespective of the sign $\pm$. 

Suppose that $A^{(0)}$ is a solution of equation of motion and
$F^{(0)}$ is the corresponding field strength. We denote the fluctuations
around this solution  $\delta A$, $A= A^{(0)} + \delta A$. 
Let us expand the left hand side of Eq.~(\ref{eqn:bogomolnyi}) 
in terms of these fluctuations: 
\begin{align}
F - {\bm{i}} \alpha \gamma_7 *_6 F \wedge F 
&= {\cal B}_0 + {\cal B}_1(\delta A) + {\cal B}_2( \delta A) ~,
\end{align}
where
\begin{align}
{\cal B}_0
 :=&  F^{(0)} -  {\bm{i}} \alpha \gamma_7 *_6 F^{(0)}  \wedge F^{(0)}  ~,\cr
{\cal B}_1 ( \delta A )
 :=&D_0 \delta A - {\bm{i}} \alpha \gamma_7 *_6  
\left(D_0  \delta A \wedge F_0  + F_0 \wedge D_0 \delta A \right) ~, \cr
{\cal B}_2 ( \delta A ) 
 :=& {\bm{q}} \delta A \wedge \delta A  - {\bm{i}} \alpha \gamma_7 *_6 
 \left( {\bm{q}} \delta A \wedge \delta A  \wedge F_0 
 + F_0 \wedge  {\bm{q}} \delta A \wedge \delta A  \right)  \cr
&- {\bm{i}} \alpha \gamma_7 *_6  \left[  \left( D_0 \delta A
 + {\bm{q}} \delta A \wedge \delta A   \right) \wedge
 \left( D_0 \delta A + {\bm{q}} \delta A \wedge \delta A   \right) \right] ~.
\end{align}
Here ${\cal B}_0$ is the zero-th order term with respect to $\delta A$. 
The term ${\cal B}_1 ( \delta A )$ is linear in $\delta A$. 
 The remaining ${\cal B}_2 ( \delta A )$ includes higher order terms. 
By substituting this to Eq.~(\ref{eqn:bogom}), we obtain
\begin{eqnarray}
&&-\frac{1}{16} \int {\rm Tr} \left\{ -  F \wedge * F
 +  \alpha^2 (F \wedge F) \wedge * (F \wedge F) \right\}  \cr
&&\quad=- \frac{1}{16} \int dv^{(4)} {\rm Tr} 
\left\{  {\cal B}_0  \wedge *_6   {\cal B}_0 
+ 2 {\cal B}_0 \wedge *_6 {\cal B}_2 ( \delta A )  \right\}  \cr
&&\qquad
- \frac{1}{16} \int dv^{(4)} {\rm Tr} 
\left\{  {\cal B}_1(\delta A)  \wedge *_6 {\cal B}_1(\delta A) \right\}
 + O(\delta A^3 ) + \int (\mbox{total derivative})~.
\label{eqn:bogom2}
\end{eqnarray}
Here the term ${\cal B}_0  \wedge *_6 {\cal B}_1 ( \delta A )  $ 
is a total derivative term because $A^{(0)}$ is a solution of the equation
 of motion.  The term $2 {\cal B}_0 \wedge *_6 {\cal B}_2 ( \delta A )$  
includes indefinite quadratic form of $\delta A$, which might yield 
a tachyonic mass term. 
When $A^{(0)}$ is a solution of ${\cal B}_0=0$ which is one 
of Eq.~(\ref{eqn:bogomolnyi}), 
no tachyonic mass term appears in gauge sector.
We mention that this does not necessarily mean the stability of 
the system under the presence of fluctuations of both the metric and 
the gauge field. This is an issue to be studied in the future.

\section{Ansatz for the metric and gauge fields}
\label{section:FCSI}
In this section we consider time-dependent solutions 
in the sense of Freund \cite{Freund:1982pg}.
Namely, the metric is assumed to be in the form,
\begin{align}
ds^2 &= ds_4^2+ds_6^2\,;
\cr
&ds_4^2=-dt^2+L_0^2e^{2\phi_1}
\frac{| d\sigma |^2}{(1+\kappa | \sigma |^2/4)^2}~,
\quad
ds_6^2= L_0^2e^{2\phi_2}\frac{ | dy |^2}{(1+| y |^2/4)^2}~,
\end{align}
where the coordinates $\sigma^i= ( \sigma^1,\sigma^2 , \sigma^3 )$ span
the three-dimensional space and $y^{I}= (y^4,y^5, \cdots , y^9)$ span $S^6$.
$| \sigma |^2 := (\sigma^1 )^2 +  (\sigma^2 )^2 +  (\sigma^3 )^2 $ and 
$| y |^2 := ( y^4 )^2 +  ( y^5 )^2 + \cdots + (y^9 )^2 $. 
The parameter $\kappa$ is $\pm 1$ or $0$. 
$\phi_1$ and $\phi_2$ are functions of time $t$. 
$L_0$ is a constant with dimension of length.  
The radius of $S^6$ is given by $R= L_0 e^{\phi_2}$.
This type of metrics was considered in various contexts,
 for instance in \cite{Ishihara:1984wx,Nariai:1986bg}.

The SO(6) gauge field configuration, represented in terms of
differential forms, is assumed to be in the form,
\begin{align}
A = \frac{1}{4 {\bm{q}}L_0e^{\phi_2}}\gamma_{ab}\,y^{a+3}V^{b+3}~,
\end{align}
where $a,b=1,2,\cdots,6$ are the indices of the Lie algebra
of SO(6), $\gamma_{ab}:=(1/2)[\gamma_a , \gamma_b ]$ 
are the infinitesimal generators represented by spinor, 
and $V^{I}$ is the vielbein of the six dimensional metric $ds_6^2$,
\begin{align}
 V^{I} &:=  L_0 e^{\phi_2}  \frac{dy^{I}}{(1+|y|^2/4)} ~.
\end{align}
${\bm{q}}$ is the gauge coupling constant.
 In the configuration of the gauge field $A$, 
the internal indices $a,b,\cdots$ and the spatial indices $I,J, \cdots$
 are identified by an embedding of the spin connection of the
 six-dimensional sphere into the gauge group.

The exterior derivatives of the vielbeins are
\begin{align}
d V^{I} &= - L_0e^{\phi_2}  
\frac{\delta_{JK} y^{J} dy^{K} \wedge dy^{I}}{2(1+|y|^2/4)^2}  + 
  L_0 \frac{\dot{\phi}_2 e^{\phi_2} dt \wedge dy^{I}}{(1+|y|^2/4)} \cr
&= - \frac{ \delta_{JK} y^{J} V^{K} \wedge V^{I} }{2L_0e^{\phi_2}} 
 +  \dot{\phi}_2 dt  \wedge V^{I} .
\end{align}
Then the Ricci tensor components are given by
\begin{align}
{\mathcal R}_{tt} &
= - 3 (\ddot{\phi}_1 + \dot{\phi}_1^2 ) - 6 (\ddot{\phi}_2 + \dot{\phi}_2^2 ) \cr
{\mathcal R}_{ij} &= 
g_{ij}  \left(  \ddot{\phi}_1   + 2 \frac{\kappa}{L_0^2 e^{2\phi_1}}  
 + 3  \dot{\phi}_1^2   + 6  \dot{\phi}_1 \dot{\phi}_2 \right)  \cr
{\mathcal R}_{IJ} &= g_{IJ}  \left( \ddot{\phi}_2  + 
5 \frac{1}{L_0^2 e^{2\phi_2}}   + 6 \dot{\phi}_2^2 
 + 3 \dot{\phi}_1 \dot{\phi}_2 \right) ~~. 
\end{align}
The scalar curvature is
\begin{align}
{\mathcal R} &= 6 \ddot{\phi}_1  +  12 \ddot{\phi}_2 
  + 12 \dot{\phi}_1^2  + 42 \dot{\phi}_2^2 
  + 36  \dot{\phi}_1 \dot{\phi}_2 
 + \frac{6}{L_0^2} \left( \kappa  e^{-2\phi_1}  + 5 e^{-2\phi_2} \right)~. 
\end{align}
Thus the Einstein tensor components are given by
\begin{align}
{\mathcal G}_{tt} 
 &=   3 \dot{\phi}_1^2  + 15 \dot{\phi}_2^2 
  + 18  \dot{\phi}_1 \dot{\phi}_2 
 + \frac{3}{L_0^2} \left( \kappa  e^{-2\phi_1}  + 5 e^{-2\phi_2} \right),
      \cr
{\mathcal G}_{ij} 
 &= - g_{ij} \left( 2 \ddot{\phi}_1  +  6 \ddot{\phi}_2 
  + 3 \dot{\phi}_1^2  + 21 \dot{\phi}_2^2 
  + 12  \dot{\phi}_1 \dot{\phi}_2 
 + \frac{1}{L_0^2} \left( \kappa  e^{-2\phi_1}  + 15 e^{-2\phi_2} \right) 
  \right), \cr
{\mathcal G}_{IJ} 
 &= - g_{IJ} \left( 3 \ddot{\phi}_1  +  5 \ddot{\phi}_2 
  + 6 \dot{\phi}_1^2  + 15 \dot{\phi}_2^2 
  + 15  \dot{\phi}_1 \dot{\phi}_2 
 + \frac{1}{L_0^2} \left( 3 \kappa  e^{-2\phi_1}  + 10 e^{-2\phi_2} \right) 
  \right) .
\end{align}

As for the gauge field, its field strength is given by
\begin{align}
F  &= \frac{1}{ 4 {\bm{q}} L_0^2 e^{2\phi_2}} \gamma_{ab}
  V^{a+3} \wedge V^{b+3}~~ .
\end{align}
This satisfies the following duality relations, 
\begin{align}
* F &= - {\bm{i}} \beta  \gamma_7\,  dv^{(4)} \wedge F \wedge F ~~,&
* (F \wedge F ) 
&=  \frac{ {\bm{i}}}{ \beta} \,dv^{(4)}  \wedge \gamma_7\, F ~~, 
\label{eqn:selfdualeq}
\end{align}
where 
\begin{align}
\beta &:= \frac{{\bm{q}} L_0^2}{3} e^{2\phi_2} \,.
\label{eqn:beta}
\end{align}

The self duality relation Eq.~(\ref{eqn:selfdualeq}) becomes 
the Bogomol'nyi equation (\ref{eqn:bogomolnyi}) if $\beta = \alpha$. 
In this case, there are no tachyonic modes at least in gauge sector. 
This determines a particular radius $R=L_c$ of the extra dimensions
in terms of the gauge coupling constants $\bm{q}$ and $\alpha$,
\begin{align}
L_c :=  \sqrt{ \frac{ 3 \alpha }{ {\bm{q}}} }~. 
\label{Lcdef}
\end{align}
Because $\phi_2$ depends only on the time coordinate,
 the exterior derivative of $\beta d^{(4)}v$ vanishes, 
\begin{align}
d( \beta d^{(4)}v  ) &= 0  ~~ . 
\end{align}
This means that the configuration satisfies the equation of motion, 
\begin{align}
D( *F ) - \alpha^2 D\{ * (F \wedge F) \wedge F 
 + F \wedge *(F \wedge F) \} =& 0 ~~. 
\end{align}

The energy momentum tensor of the gauge field is given by
\begin{align}
{\mathcal T}_{MN} 
&=  \frac{1}{8} {\rm Tr}  \left( - F_{MP} F_N{}^P 
 + \frac{\alpha^2}{3!}   H_{MPQS} H_N{}^{PQS}   \right)
 - \frac{1}{2} g_{MN} \chi  ~~, \cr
\chi &:= \frac{1}{8}{\rm Tr} \left(  - \frac{1}{2}  F_{MN} F^{MN}
 +  \frac{\alpha^2}{4!} H_{MNPQ} H^{MNPQ} + V_0 \right) ~~ . 
\end{align}
Here $H_{IJKL}$ are the components of $F \wedge F$ introduced in Appendix A. 
For our gauge configuration we have
\begin{eqnarray}
{\mathcal T}_{tt} 
&=&   \frac{1}{2} \chi ~,\quad
{\mathcal T}_{ij} 
=  - \frac{1}{2} g_{ij} \chi ~, 
\cr
{\mathcal T}_{IJ} 
 &=& - \frac{5}{8{\bm{q}}^2 L_0^4 }e^{-4 \phi_2}
 \left(    1 - \frac{ 3^2 \alpha^2}{{\bm{q}}^2 L_0^4} e^{-4 \phi_2} \right)
 g_{IJ} -  \frac{1}{2} g_{IJ} V_0 ~,
\end{eqnarray}
where
\begin{align}
\chi\equiv\chi(\phi_2) 
&=  \frac{15}{4{\bm{q}}^2 L_0^4}e^{-4\phi_2} 
\left(    1 +  \frac{3^2 \alpha^2 }{ {\bm{q}}^2L_0^4 } e^{-4\phi_2}\right)+ V_0 ~.
\label{eq:chi}
\end{align}
In this gauge configuration, the Einstein field equations are
\begin{eqnarray}
&& \frac{8 \pi G }{2} \chi 
= 3 \dot{\phi}_1^2  + 15 \dot{\phi}_2^2 
  + 18  \dot{\phi}_1 \dot{\phi}_2 
 + \frac{3}{L_0^2} \left( \kappa  e^{-2\phi_1}  + 5 e^{-2\phi_2} \right)
 \,,
\label{eqn:EE1}\\
&&\frac{8 \pi G }{2} \chi  
= \left( 2 \ddot{\phi}_1  +  6 \ddot{\phi}_2 
  + 3 \dot{\phi}_1^2  + 21 \dot{\phi}_2^2 
  + 12  \dot{\phi}_1 \dot{\phi}_2 
 + \frac{1}{L_0^2} \left( \kappa  e^{-2\phi_1}  + 15 e^{-2\phi_2} \right) 
  \right)
\,,    
\label{eqn:EE2}
\\
&&8 \pi G \left[ - \frac{5}{8{\bm{q}}^2 L_0^4 }e^{-4 \phi_2}
  \left(    1 - \frac{ 3^2 \alpha^2}{{\bm{q}}^2 L_0^4} e^{-4\phi_2}   \right)
  -  \frac{1}{2}  V_0 \right] 
 \cr
&&\qquad
= - \left( 3 \ddot{\phi}_1  +  5 \ddot{\phi}_2 
  + 6 \dot{\phi}_1^2  + 15 \dot{\phi}_2^2 
  + 15  \dot{\phi}_1 \dot{\phi}_2 
 + \frac{1}{L_0^2} \left( 3 \kappa  e^{-2\phi_1}  + 10 e^{-2\phi_2} \right) 
  \right)
  \,,
\label{eqn:EE3}
\end{eqnarray}
where the first equation is a constraint on the field and its derivatives,
the Hamiltonian constraint equation,
determining the three-dimensional hypersurface in the four-dimensional
phase space.
We note that the kinetic term in the Hamiltonian constraint is quadratic
in the field velocities, and it has one positive and one negative
eigenvalues.
The above system of differential equations is invariant under 
the time translation and time reversal transformation. If $\kappa=0$,
there is in addition an invariance under the shift of $\phi_1$.
The time evolution of the fields $\phi_1$ and $\phi_2$ is 
determined by Eqs.~(\ref{eqn:EE2}) and (\ref{eqn:EE3}),
describing the trajectory on the three-dimensional  
hypersurface defined by the constraint equation.

It is convenient to express the field
equations in terms of a rescaled time 
coordinate $\tau=t/L_0$,
and introduce the following dimensionless 
parameters,
\begin{align}
a&=\frac{8\pi G}{{\bm{q}}^2 L_0^2}, \\
b&=\frac{\alpha^2}{{\bm{q}}^2L_0^4}, \\
c&=4\pi G V_0L_0^2, 
\end{align}
where $c$ is related to the ten-dimensional cosmological 
constant $\Lambda$ by $c=\Lambda L_0^2$.
In what follow, the $\tau$-derivative of a function $h(\tau)$
will be denoted by $h'$.
For $L_0=L_c$ the parameter $b$ is fixed to the
value $b=1/9$, leaving only two free parameters
in the field equations.

By manipulating the field equations we
can reduce them to the following convenient set of
two differential equations:
\begin{align}
V_1 &=  (\phi_1')^2  + 
5(\phi_2')^2   + 6  (\phi_1') (\phi_2')  \label{eqn:EEas1}\,,  \\
V_2&={\phi}''_2+6(\phi_2')^2+
3(\phi_1') (\phi_2') \label{eqn:EEas3}
\,,  
\end{align} 
where $V_1$ and $V_2$ are defined by
\begin{align}
V_1(\phi_1,\phi_2) &:=\frac{4\pi GL_0^2}{3}
\,\chi\left(\phi_2\right)
-5 e^{-2\phi_2}
-\kappa e^{-2\phi_1}
\,, \\
&= \frac{5}{8}a e^{-4 \phi_2} \left( 1 + 9 b e^{-4 \phi_2}  \right) 
+ \frac{c}{3} - 5 e^{-2 \phi_2} - \kappa e^{-2 \phi_1} ~,\cr
V_2(\phi_2)&:=
\frac{5a}{32}e^{-4 \phi_2}  \left(5+63 b e^{-4\phi_2}\right)  
+\frac{c}{4}-
5 e^{-2\phi_2}
\,.
\end{align}

We can solve Eq.~(\ref{eqn:EEas1}) for $\phi_1'$
to obtain
\begin{align}
\phi_1' =- 3 \phi_2' \pm \sqrt{  V_1 + 4 ( \phi_2')^2 }
\,.
\end{align}
By using this equation,
we can eliminate $\phi_1'$ from Eq.~(\ref{eqn:EEas3}).
Then the Einstein field equations are reduced to
a system of coupled differential 
equations given by
\begin{align}
\label{eqn:diffphi_1}
\phi_1' + 3 \phi_2' -\sqrt{  V_1 + 4 ( \phi_2')^2 }
&=0
\,,\\
\label{eqn:ddphi_2}
\phi_2'' -3 (\phi_2')^2+
 3\phi_2'\sqrt{  V_1 + 4 ( \phi_2')^2 }
-V_2 &=0
\,,
\end{align}
where we have chosen the positive value of the square root
in Eq.~(\ref{eqn:diffphi_1}). In the next section, we look for
a solution in which the extra-dimensional part of the metric is static,
that is, a solution with $\phi_2'=0$. In this case, $\phi_1$ grows with
time for the above choice of the square root sign,
ensuring that the four-dimensional part of the metric
describes an expanding universe.

\section{Solutions with static extra dimensions}
\label{section:simple}

In this section we consider solutions in which the metric 
of the extra dimensional space, $S^6$, is static, that is
when $\phi_2=$constant.
In this case, Eq.~(\ref{eqn:diffphi_1}) becomes integrable
with respect to $\phi_1$, and Eq.~(\ref{eqn:ddphi_2}) becomes
an algebraic equation for $e^{-2\phi_2}$.
We note that we do {\it not\/} require our solution to satisfy the
Bogomol'nyi equation (\ref{eqn:bogomolnyi}). Hence for those solutions
whose extra-dimensional radius is different from $L_c$ given by
 Eq.~(\ref{Lcdef}), the absence of tachyon modes is not guaranteed.
Therefore we simply assume that there is a sufficiently wide range
of parameters in which there appears no harmful tachyons.
This issue is left for a future study.

Below we first consider general solutions.
As we will see shortly, there is a particular solution
given by $ e^{\phi_1}=\sqrt{-\kappa}\tau+C$ for $\kappa=-1,0$. 
Since this is somehow special, we treat it separately.

\subsection{The general case}

Static solutions of Eq.~(\ref{eqn:ddphi_2}) are determined by 
the roots of $V_2(\phi_2)=0$. Let us set $Z:= a e^{-2 \phi_2}$.
We note that $Z\propto(L_0e^{\phi_2})^{-2}$, where $L_0e^{\phi_2}$
is the linear scale of the extra dimensions. 
The equation $V_2=0$ becomes
\begin{align}
f(Z)\equiv  Z^4 + 5  \nu_1  Z^2 - 32 \nu_1 Z   +  \frac{8\nu_2\nu_1}{5} &=0  ~,
\label{eqn:solv2}
\end{align}
where 
\begin{align}
\nu_1 &:= \frac{a^2}{63b}= \frac{64\pi^2 G^2}{63\alpha^2{\bm{q}}^2}~,&
\nu_2 &:= ac = \frac{32 \pi^2 G^2 V_0}{{\bm{q}}^2}~.
\end{align}
Note that $\nu_1$ and $\nu_2$ are independent of $L_0$ and $\phi_2$.
As demonstrated in Appendix~\ref{sec:cals}, the equation $f(Z)=0$ has
one or two real solutions $Z_1$ and $Z_2$ (we assume $Z_1\geq Z_2$) 
provided that $\nu_1$ and $\nu_2$ satisfy a certain inequality.

Let us first consider the solution $Z_1$. 
The relation between the original variables and $Z_1$ can be written as 
\begin{equation}
L_0^2\exp(2 \phi_2)=L_1^2:=\frac{8\pi G}{\bm{q}^2}\frac{1}{Z_1},
\label{L1def}
\end{equation}
where $L_1$ represents the size of the extra dimensions. 
Thus the size of the extra dimensions is completely fixed by 
the coupling constants.

As easily seen, Eqs.~(\ref{eqn:diffphi_1}) and (\ref{eqn:ddphi_2}) are
invariant under the rescaling,
\begin{equation}
L_0\rightarrow C L_0~~,~~\exp(\phi_2)\rightarrow C^{-1}\exp(\phi_2)
~~,~~\exp(\phi_1)\rightarrow C^{-1}\exp(\phi_1). 
\end{equation}
Using this degree of freedom, we fix the length scale $L_0$ 
to be the size of the extra dimensions $L_1$, or equivalently,
we set $\phi_2=\phi_2^{(1)}=0$ for this solution.
Then we have $Z_1=a$. Therefore $a$ must be a solution of
Eq.~(\ref{eqn:solv2}):
\begin{equation}
f(a)=0\Leftrightarrow c=20-\frac{5a}{8}(5+63b)\,.
\label{eq:ctoab}
\end{equation}
With this normalization, we find a positive real $Z_2$ for $c>0$.
The condition $Z_2 \leq Z_1$  and $c\geq 0$ give the following inequalities:
\begin{align}
\frac{ 32}{5+63b} \geq a &\geq \frac{16}{5+126 b} ~.
\label{eqn:bounda}
\end{align}

\begin{figure}[htbp]
\begin{center}
\includegraphics[scale=2]{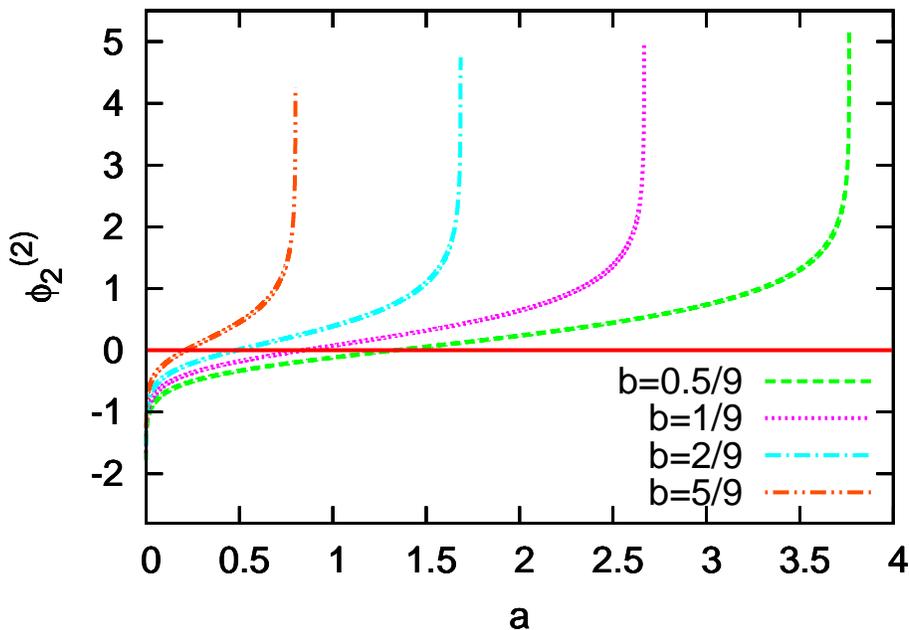}
\caption{Plot of $\phi_2^{(2)}$ as a function of $a$ 
for $b= 0.5/9$, $1/9$, $2/9$ and $5/9$. 
 Because $\phi_2^{(2)} \geq \phi_2^{(1)}=0$, only the part
of the curves above the line $\phi_2^{(2)}=0$ is meaningful.
The value of $a$ is bounded as given by Eq.~(\ref{eqn:bounda}).}
 \label{fig:phi2}
\end{center}
\end{figure}

In terms of $\phi_2$ these two solutions are given by
\begin{align}
\phi_2^{(1)} &:= - \frac{1}{2} \log ( Z_1/a ) = 0~,& 
\phi_2^{(2)} &:= - \frac{1}{2} \log ( Z_2/a )~.
\end{align}
Those points are critical points  or equilibrium solutions of
the differential equation~(\ref{eqn:ddphi_2}).
The value of $\phi^{(2)}_2$ is depicted as a function of $a$ 
for each value of $b$ in Fig.~\ref{fig:phi2}. 
The discussions in the rest of this subsection is valid for
both solutions.

\begin{figure}[htbp]
\begin{center}
\includegraphics[scale=1]{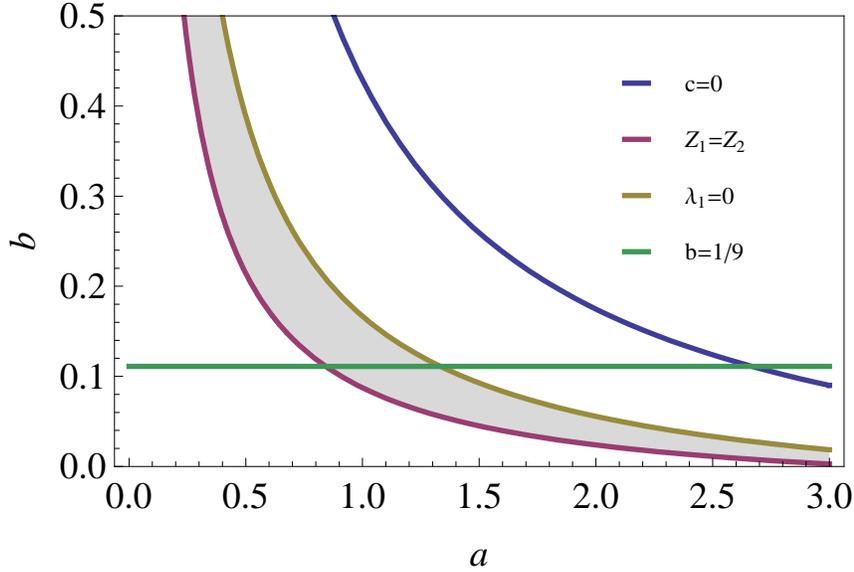}
\caption{Allowed region of $a$ and $b$. 
The filled region is the allowed region 
which is bounded by the lines $\lambda_1^2=0$ and $c=0$. 
}
\label{fig:ab}
\end{center}
\end{figure}

Now we turn to Eq.~(\ref{eqn:diffphi_1}).
Setting $\phi_2'=0$, we have
\begin{eqnarray}
\phi'_1 = \sqrt{\lambda_i^2-\kappa e^{-2\phi_1}}
\quad\Leftrightarrow\quad
(e^{\phi_1})'=\sqrt{\lambda_i^2e^{2\phi_1}-\kappa}\,,
\label{eqn:odephi2}
\end{eqnarray}
where $\lambda_i$ ($i=1,2$) is defined by
\begin{align}
\lambda_i^2 :=\frac{4}{3}\pi G L_0^2\,\chi(\phi^{(i)}_2)
 -5\,e^{-2\phi^{(i)}_2}\,.
\label{eqn:lambda}
\end{align}
We assume $\lambda_i^2$ is positive. 
  $\lambda_1^2 \geq 0$ gives an additional condition on the parameter $a$, 
\begin{align}
 \frac{4}{1+18 b} \geq a  \geq \frac{16}{5+126 b} ~. 
\label{eqn:bounda2}
\end{align}
If this inequality is satisfied, $\lambda_2^2 \geq 0$, because  
$\frac{4}{3}\pi G L_0^2 \chi(\phi_2)- 5 e^{- 2\phi_2}$ is concave 
downward as a function of $e^{-2 \phi_2}$ and its derivative at
$e^{-2\phi_2}=1$ is negative. The allowed region of $a$ and $b$ given
by Eq.~ (\ref{eqn:bounda2})  is depicted in Fig.~\ref{fig:ab}. 
%
%

The equation~(\ref{eqn:odephi2}) can be integrated to give 
\begin{align}
e^{\phi_1} &= \frac{1}{2 \lambda_i} 
\left(e^{\lambda_i \tau} + \kappa e^{- \lambda_i \tau} \right) ~ ,
\label{phi1sol}
\end{align}
where the origin of the time coordinate has been chosen to
make the expression simple.
Four-dimensional parts of these solutions are the
same as those of Ishihara  \cite{Ishihara:1984wx}. 

For large $\tau$, the term proportional to $\kappa$ can be neglected
and the scale factor of the four-dimensional space-time approaches
$R(t)=L_0e^{\lambda_i \tau} $, which describes a universe
with accelerated expansion.
Thus, although we do not claim that our model can give a realistic model
of the universe, depending on the value of the constant $\lambda_i$,
it can reproduce a period of inflation in the very early universe
or the present universe dominated by a very small cosmological constant.

\subsection{The case $\lambda_i=0$}

When $\lambda_i=0$ the solution (\ref{phi1sol}) is no longer
valid as it is, and we need a special treatment.
In this case Eq.~(\ref{eqn:odephi2}) implies that
$\kappa$ must be either $-1$ or $0$. In either case, 
$(e^{\phi_1})'=\sqrt{-\kappa}$, and the solution is
\begin{equation}
e^{\phi_1}=\sqrt{-\kappa}\tau +C
\quad(\kappa=-1,0)\,,
\label{phi1flatsol}
\end{equation}
where $C$ is an integration constant. 

The four-dimensional part of the solution for $\kappa=0$ is flat.
It was obtained in \cite{Kihara:2007vz}, 
which is almost the same as the one obtained by
 Cremmer-Scherk \cite{Cremmer:1976ir}, but with the radius of $S^6$
and the value of ten-dimensional cosmological constant modified by
the presence of the Tchrakian term. 

The solution for $\kappa=-1$ is also flat.
The four dimensional line element is
\begin{equation}
ds^2
= L_0^2 \left( -d\tau^2+\tau^2\frac{d\sigma^2}{(1-\frac{|\sigma|^2}{4})^2}
 \right)
\,.
\end{equation}
This metric covers the inside of either the future light cone
or the past light cone of the flat space-time.

\section{Dynamical Compactification}
\label{section:dynamical}

In this section, we switch on the time dependence of $\phi_2$
in order to see if our model has the possibility to describe 
the process of dynamical compactification.
For this purpose, we analyze the stability of the
 solution $\phi_2=\phi_2^{(1)}$ ($=0$) and $\phi_2=\phi_2^{(2)}$
in the second order nonlinear differential equation (\ref{eqn:ddphi_2})
in the case of $\kappa=0$ 
both analytically and numerically.

We first analyze the stability of the critical points
analytically. For this purpose, we linearize the system of 
differential equations (see e.g. \cite{ODE}). We find, however, 
that for $\lambda_1=0$ this method is not sufficient to establish the 
stability of the critical point $\phi_2'=\phi_2=0$. Therefore we 
will try a different approach in this case.

We first consider the critical point $(\phi_2,\phi_2')=(0,0)$,
which is a stationary or equilibrium
solution of the differential equation (\ref{eqn:ddphi_2}).
Denoting $X:=\phi_2$ and $Y:=\phi_2'$ and keeping only terms
linear in $X$ and $Y$, Eq.~(\ref{eqn:ddphi_2}) is
written as the following system of first order differential equations:
\begin{align}
\mathbf X':=\left(
\begin{array}{c}
X'\\
Y'
\end{array}
\right)
=
\left(
\begin{array}{c}
Y \\
\left.\frac{dV_2}{d\phi_2}\right|_{\phi_2=0}X-3\lambda_1 Y
\end{array}
\right)=\mathbf A \mathbf X\,,
\label{eqn:sys}
\end{align} 
where the matrix $\mathbf A$ is given by 
\begin{align}
\mathbf A=
\left(
\begin{array}{cc}
0&1\\
\left.\frac{dV_2}{d\phi_2}\right|_{\phi_2=0}
&~~
-3\lambda_1~~
\end{array}
\right)
=\left(
\begin{array}{cc}
0&1\\
-\Omega^2+\frac{21}{2}\lambda_1^2
&~~
-3\lambda_1~~
\end{array}
\right)
\end{align}
with 
\begin{align}
\lambda_1^2 &= \frac{5}{3} - \frac{5}{12} a ( 1 + 18 b)
\,, \\ 
\Omega^2 &:= \frac{5}{4} (6-a) ~.
\end{align}
Note that $\lambda_1$ in the above is equal to the one defined by
Eq.~(\ref{eqn:lambda}) with the normalization condition~(\ref{eq:ctoab}).
The solution $(X,Y)=(0,0)$ is asymptotically stable
if both of the two eigenvalues of the matrix $\mathbf A$,
\begin{equation}
\Upsilon_{1,\pm}=-\frac{3}{2}
\left[\lambda_1\pm
\sqrt{\lambda_1^2+\left.{\frac{4}{9}\frac{dV_2}{d\phi_2}}\right|_{\phi_2=0}}\,
\right]
\,,
\end{equation}  
have negative real part. 

To analyze the stability of the second critical point,
$(\phi_2,\phi_2')=(\phi_2^{(2)},0)$, we simply replace $\phi_2$ by
$\phi_2-\phi_2^{(2)}$ when linearizing Eq.~(\ref{eqn:ddphi_2}).
Then the eigenvalues are
\begin{align}
\Upsilon_{2,\pm}=-\frac{3}{2}\left[\lambda_2\pm
\sqrt{\lambda_2^2+\frac{4}{9}
\left.\frac{dV_2}{d\phi_2}\right|_{\phi_2=\phi^{(2)}_2}}
\right]
\,.
\end{align}

For $\lambda_1=0$ the real part of the two 
eigenvalues is zero, and the linear system 
corresponds to the harmonic oscillator for
$\Omega^2>0$. 
In this case we can not apply Poincare-Lyapunov's
theorem above and additional information is required 
to establish the character of the critical point 
for the full nonlinear equation. We therefore treat 
this case separately.

\subsection{The case $\lambda_1>0$}
                 
As in Sec.~\ref{section:simple},
we are interested in the solutions with
$\lambda_1$ real and positive.
Then the real part of the eigenvalues $\Upsilon_{1,\pm}$
is negative if
\begin{align}
-\left.\frac{dV_2}{d\phi_2}\right|_{\phi_2=0}
=\left(\Omega^2-\frac{21}{2}\lambda_1^2\right)>0
\,.
\end{align}
This condition coincides with the condition $Z_2<Z_1$,
which is satisfied when the parameters satisfy
Eq.~(\ref{eqn:bounda2}). Thus the critical point
$(\phi_2,\phi_2')=(0,0)$ is stable.

The system shows two different kinds of behavior in the
neighborhood of the critical point $(\phi_2,\phi_2')=(0,0)$.
When $\lambda_1^2-\frac{4}{9}\left(\Omega^2-\frac{21}{2}\lambda_1^2\right)<0$,
the system undergoes damped oscillations with the amplitude
decreasing as $e^{-3/2\lambda_1 \tau}$. 
Otherwise the system is over-damped, showing
simple exponential damping toward the critical point.


For the second critical point $(\phi_2,\phi_2')=(\phi_2^{(2)},0)$, 
it can be shown that 
\begin{align}
\left. \frac{dV_2 }{d \phi_2} \right|_{\phi_2^{(2)}} 
&= - \frac{5}{16} \frac{63b}{a^3} Z_2 \frac{df}{dZ}(Z_2) \geq 0 ~,
\end{align}
where $f(Z)$ is the function introduced in Eq.~(\ref{eqn:solv2})
and $Z_2$ is the solution of $f(Z)=0$ corresponding to the
second critical point. This inequality follows from the fact that
$df/dZ$ is a monotonically increasing function of $Z$ with
the unique $df/dZ=0$ at $Z=Z_0$ and $Z_2\leq Z_0$, which is proved
in Appendix~\ref{sec:cals}.
If $df/dZ(Z_2) <0$, there are two real eigenvalues with opposite signs.
Hence the critical point is an unstable saddle-point.
In the special case when $df/dZ(Z_2)=0$, the first and second
critical points become degenerate, and $\phi_2=0$ becomes the only
equilibrium solution of the system. Note that we have
$\Omega^2=21/2\lambda_1^2$ in this case.

\begin{figure}[htbp]
\begin{center}
\includegraphics[width=0.8\textwidth]{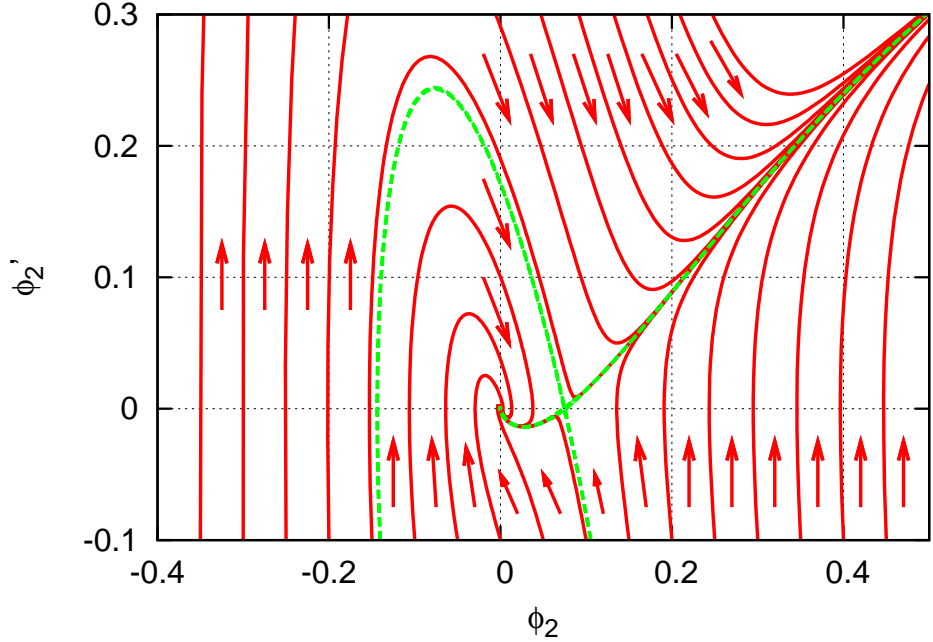}
\end{center}
\caption{\scriptsize Phase space diagram $(\phi_2,\phi_2')$ 
for $\lambda_1^2=5/12$ and $\Omega^2=25/4$. These are equivalent 
to take $a=1$ and $b=1/9$. The figure shows the two critical
points of the system.
The point (0,0)is stable while 
the second critical point is a saddle point with unstable 
orbits to its right.}
\label{fig:phase_diagr}
\end{figure}

To confirm the above stability analysis, we have
performed numerical integration of Eq.~(\ref{eqn:ddphi_2}).
Our numerical results indicate that the linear analysis 
around the first critical point is accurate.
In Fig.~\ref{fig:phase_diagr}, we show the phase-space orbits
of the solutions of Eq.~(\ref{eqn:ddphi_2}) with $\lambda_1>0$.
In this case, the first critical point is stable and the other
critical point along the $\phi_2'=0$ axis is an unstable saddle point. 
The location of the saddle point depends on the 
values of the parameters $a$ and $b$ as well, and it 
roughly defines an effective stability radius for orbits 
near the solution $(0,0)$. The time evolution of $\phi_2$
for an asymptotically stable solution is shown in Fig.~\ref{fig:oscl1},
where $ e^{\phi_2}$ oscillates with a decreasing amplitude until
$\phi_2$ reaches zero.  

\begin{figure}[htbp]
\begin{center}
\includegraphics[scale=2]{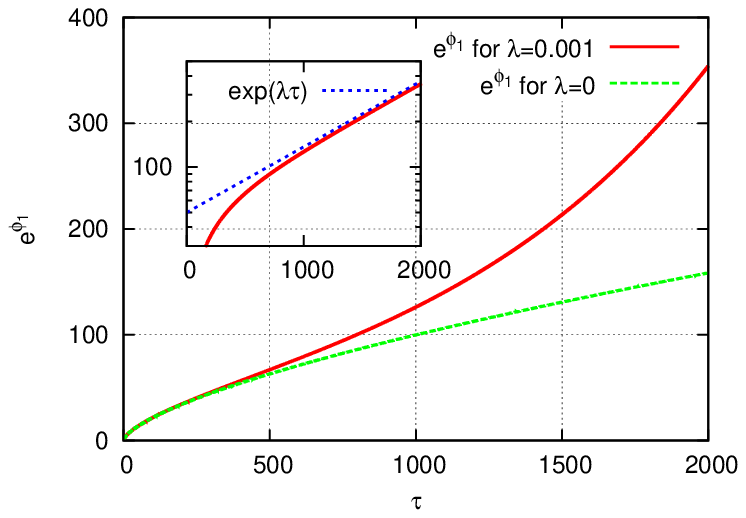}
\caption{\scriptsize This figure shows the time evolution
of the scale factor $R(\tau)$ for $\Omega^2=25/4$ and 
$\lambda_1 =0.001$. 
}
\label{fig:oscln2}
\end{center}
\end{figure}

We have also integrated Eq.~(\ref{eqn:diffphi_1}) for $\phi_1$.
The time evolution of the three-dimensional cosmic scale factor
$R\propto e^{\phi_1}$ is shown in Fig.~\ref{fig:oscln2}.
Initially when the oscillatory energy of $\phi_2$ is non-negligible,
the scale factor behaves as the one in a matter-dominated universe,
$R(\tau)\propto\tau^{2/3}$. For sufficiently large $\tau$,
after the amplitude of $\phi_2$ has decayed exponentially,
the universe eventually enters a stage of accelerated expansion,
$R(\tau)\propto e^{H\tau}$, with the (dimensionless) 
Hubble parameter $H=\lambda_1$.

\begin{figure}[htbp]
\begin{center}
\includegraphics[scale=2]{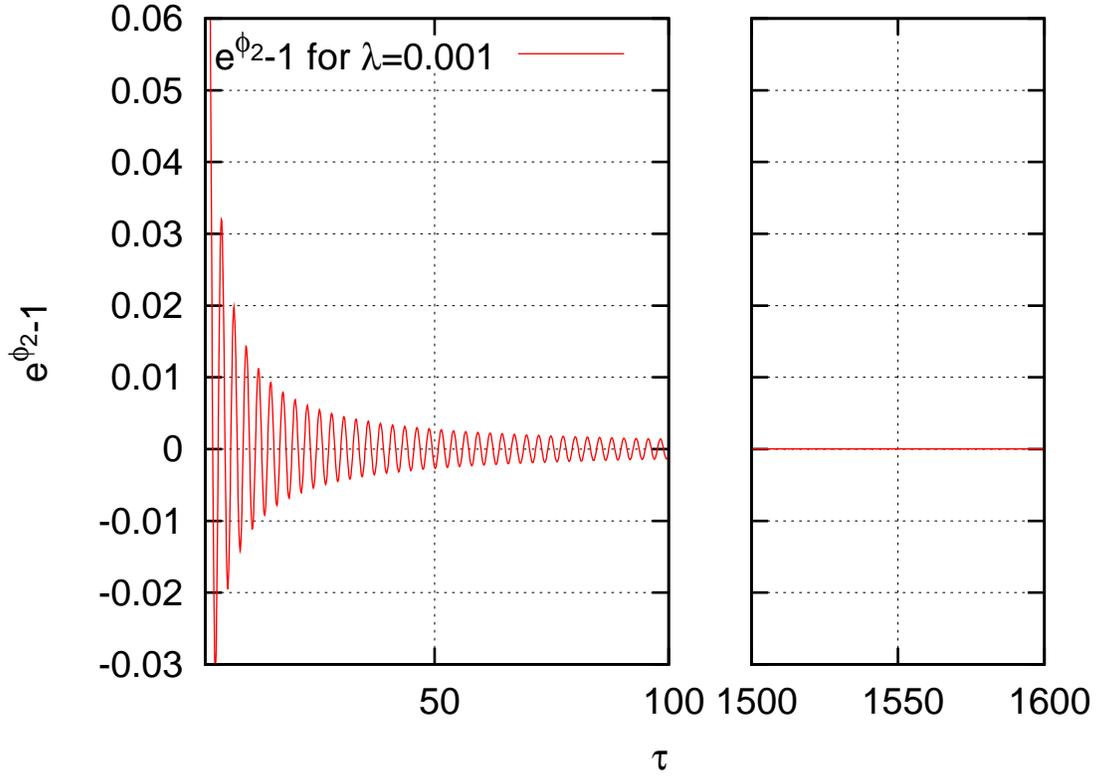}
\caption{\scriptsize This figure shows the damped 
oscillations of the radius of the extra dimensions 
with time $\tau=tL_0^{-1}$. 
}
\label{fig:oscln1}
\end{center}
\end{figure}

\subsection{The case $\lambda_1=0$}

In the case $\lambda_1=0$ the real part of $\Upsilon_{1,\pm}$ is zero,
rendering the linear analysis insufficient to determine 
the stability of the solution. Therefore we have to take into account
the second order terms. 

To second order in $X$ and $Y$, Eq.~(\ref{eqn:ddphi_2}) gives
the equations,
\begin{align}
X'&=Y \label{eqn:sys3}
\,,\\
Y'&=-\Omega^2X+G\left(X,Y\right)
\label{eqn:sys4}
\,.
\end{align}
where $G(X,Y)$ is a quadratic function 
given by 
\begin{align}
G\left(X,Y\right)=\left(9\Omega^2-\frac{15}{2}\right)X^2
+3Y^2-3Y\sqrt{4\Omega^2X^2+4Y^2}
\,.
\end{align}

Let us solve Eqs.~(\ref{eqn:sys3}) and (\ref{eqn:sys4})
perturbatively. We assume $\Omega^2>0$.
To first order in $X$ and $Y$, the system describes a harmonic oscillator.
Namely we have 
\begin{align}
X(\tau)&=r\cos(\Omega\tau+\psi),\\
Y(\tau)&=-\Omega\, r\sin(\Omega\tau+\psi)
\end{align}
as a solution of the first order equations, 
where $r$ and $\psi$ are arbitrary constants. 
Then the  orbits in phase-space 
are ellipses about the critical point 
$(0,0)$.

Now we consider the effect of the second order terms.
Here we just apply the so-called Krylov-Bogoliubov method 
of averaging \cite{ODE} to study the behavior of the 
solutions.\footnote{Detailed calculation
is shown in Appendix~\ref{sec:withoutav}.}

First, we introduce varying constants in
the harmonic oscillator solution as 
\begin{align}
X(\tau)&=r(\tau)\cos\left(\Omega\tau+\psi(\tau)\right)
\,, 
\label{eqn:Xtau}
\\
Y(\tau)&=-\Omega r(\tau)\sin\left(\Omega\tau+\psi(\tau)\right). 
\label{eqn:Ytau}
\end{align}
Then the system of differential equations
may be expressed as
\begin{align}
r'&=f_r(\tau,r,\psi)
\,,
\label{eqn:r}\\
\psi'&=f_\psi(\tau,r,\psi)
\,, 
\label{eqn:psi}
\end{align}
where 
\begin{align}
f_r(\tau,r,\psi)=&
-\frac{1}{\Omega}
\sin\left(\Omega\tau+\psi\right)
G\left(r\cos\left(\Omega\tau+\psi\right),
-\Omega r\sin\left(\Omega\tau+\psi\right)\right)
\,,\\
f_\psi(\tau,r,\psi)=&
-\frac{1}{\Omega r}
\cos\left(\Omega\tau+\psi\right)
G\left(r\cos\left(\Omega\tau+\psi\right),
-\Omega r\sin\left(\Omega\tau+\psi\right)\right)
\,. 
\end{align}
Note that the right-hand sides of Eqs.~(\ref{eqn:r})
and (\ref{eqn:psi}) are periodic in $\tau$ with
the period $2\pi\Omega^{-1}$. Then
instead of these equations, applying
the Krylov-Bogoliubov method of averaging
we consider the time-averaged equations: 
\begin{align}
\bar r'&=\frac{\Omega}{2\pi}\int^{2\pi/\Omega}_0
f_r(s,\bar r,\bar \psi)ds=-3\Omega \bar r^2, \\
\bar \psi'&=\frac{\Omega}{2\pi}
\int^{2\pi/\Omega}_0
f_\psi(s,\bar r,\bar \psi)ds=0 
\end{align}
for $\bar r$ and $\bar \psi$. 
The solution is given by 
\begin{align}
\bar r&=\frac{1}{3\Omega \tau+{\rm const.}},\\
\bar \psi&={\rm const.}. 
\end{align}
These give approximate behavior of $r$ and $\phi$
at sufficiently large $\tau$. From Eq.~(\ref{eqn:Xtau}), 
approximations to $\phi_2$ and $\phi_2'$ for large $\tau$ 
are given by
\begin{align}
\phi_2\left(\tau\right)&\sim
\frac{1}{3\Omega\tau}
\cos\Omega\tau
\,, \\
\phi_2'(\tau)&\sim
-\frac{1}{3\tau}\sin\Omega\tau
\,.
\end{align}

Then, for large $\tau$, Eq.~(\ref{eqn:diffphi_1}) gives
\begin{align}
\phi_1'\sim-\frac{\sin\Omega\tau}{\tau}+\frac{2}{3\tau}. 
\end{align}
We can now read off an approximate solution 
for the field $\phi_1$,
\begin{align}
\phi_1 &\ \sim \frac{2}{3} \log \tau - {\rm Si}(\Omega \tau )~.
\label{eqn:phi1}
\end{align}
Thus the scale factor behaves as
\begin{align}
R(\tau)=L_0e^{\phi_1(\tau)}\sim L_0 \tau^{2/3} e^{-{\rm Si}(\Omega \tau)}\,.
\end{align}
Apart from the small oscillations, this describes a matter-dominated
universe.

\begin{figure}[htbp]
\begin{center}
\includegraphics[scale=2]{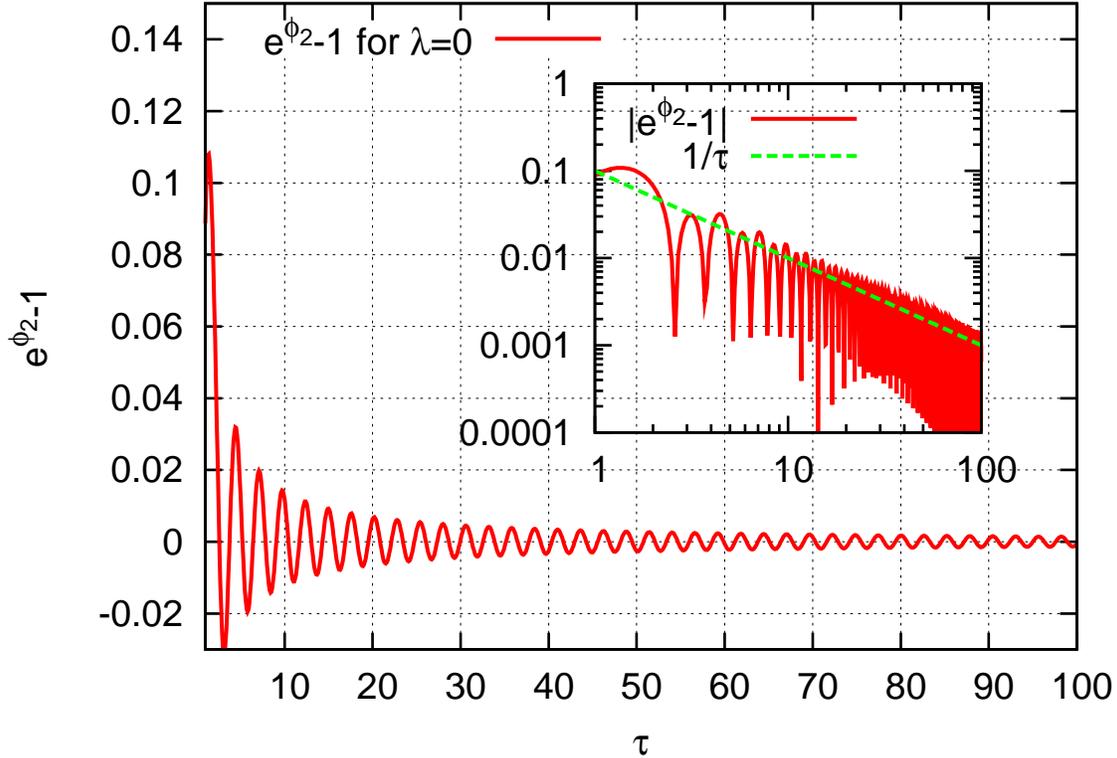}
\caption{\scriptsize The time evolution of 
$( e^{\phi_2}-1)/L_0$  with time $\tau=t/L_0$.
We have chosen $\Omega=25/4$, corresponding to
the choice of $a=1$.
The oscillations in the proximity of the equilibrium
solution $\phi_2=0$ are rapidly damped out as 
$1/\tau$.}
\label{fig:oscl0}
\end{center}
\end{figure}

\begin{figure}[htbp]
\begin{center}
\includegraphics[scale=2]{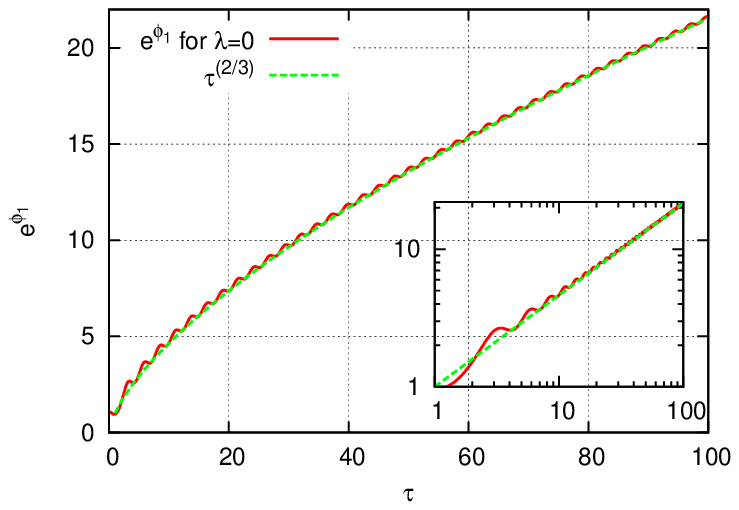}
\caption{\scriptsize The time evolution 
of $e^{\phi_1}/L_0$ for $\lambda=0$. In this plot   
$\Omega=25/4$, which corresponds to $a=1$. 
The time-averaged scale factor
$a(t)\propto< e^{\phi_1}>$ describes a matter-dominated universe.
}
\label{fig:oscl1}
\end{center}
\end{figure}

In Figs.~\ref{fig:oscl0}-\ref{fig:oscl1},
we show numerical solutions of the full
non-linear system for $\lambda_1=0$.
The numerical results are in good agreement 
with our analytical estimations. 
The time evolution of $e^{\phi_1}$ in 
Fig.~\ref{fig:oscl1} clearly exhibits oscillations
around its central value $\tau^{2/3}$ as we have 
shown analytically.


\section{Conclusion and Discussion}
\label{section:conclusion}

In this article, we studied time-dependent solutions of 
the ten-dimensional Einstein-Yang-Mills theory with the Tchrakian term.
We obtained a class of simple analytic solutions in which the extra 
dimensions are static and the scale factor of the four-dimensional
Friedmann-Lemaitre-Robertson-Walker metric behaves exponentially in time
with the rate of expansion given by constants denoted by $\lambda_i$
($i=1,2$). Thus our model admits solutions describing inflation. 

We then considered a possible dynamical compactification of
the extra dimensions by allowing them to be time-dependent.
In the case $\lambda_1>0$, we found solutions in which the scale
factor of the extra dimensions undergoes damped oscillations and 
approaches a constant value, while the four-dimensional scale factor 
approaches $e^{\lambda_1 \tau}$.
In the case of $\lambda_1=0$, 
we found numerically that the scale factor behaves
as a matter-dominated universe $R\propto \tau^{2/3}$.

Our model includes four dimensionful constants $(G,V_0, {\bm{q}}, \alpha)$.
 They define four typical length scales in our model. 
Or if we fix the Planck scale or the gravitational constant, $G$,
we are left with three dimensionless parameters.
In addition, if we require the Bogomol'nyi equation to be satisfied, 
the linear size of the extra dimensions is fixed to be 
$L_c=\sqrt{3\alpha/\bm{q}}$,
and there remains only two dimensionless parameters. 

As is shown in Sec.~\ref{section:set}, when the radius of 
the compact direction is equal to $L_c$,
there are no tachyonic mode in the gauge sector.
However, for a set of model parameters that gives a radius
substantially different from $L_c$, a tachyonic mode may appear.
To investigate when a tachyon appears and how it affects our model
is certainly an important issue.
Also for a complete analysis, in addition to fluctuations
of the gauge field, it is necessary to include
fluctuations of the metric  and cross terms between them. 
These are left for future work.

We also note that all the discussions given in this paper
applies equally to the gauge group SU(4) in place of SO(6),
because the matrices $\gamma_{ab}$ are block diagonalizable. 
Namely, if we project those matrices on the four-dimensional eigenspace
with respect to the eigenvalue $+1$  of $\gamma_7$, we obtain self-duality 
relation of SU(4) without $\gamma_7$. 
Thus all cosmological solutions obtained in this paper are also
valid for models with SU(4) gauge theory. 
Furthermore, since SU(4) is a subgroup of SU($N$),
our cosmological solutions can be embedded into the Einstein-Yang-Mills
theory with the Tchrakian term with SU($N$) gauge group.
Generalization to other gauge groups like $E_8$ or SO($N$) with $N\geq8$
remains as a future issue \cite{Colgainkihara}.

Recently some of us (HK and MN) considered the Bogomol'nyi 
equation on ${\mathbb C}{\bf P}^n$ \cite{Kihara:2008zg}. 
By using the gauge configuration on ${\mathbb C}{\bf P}^3$, 
we expect that we will be able to obtain similar cosmological 
solutions for ${\mathbb C}{\bf P}^3$ compactification instead of 
$S^6$ studied in this paper. Also, it is interesting to see
if similar cosmological solutions can be obtained for other types
of compactification such as the compactification in terms of
the Casimir energy \cite{Carroll:2009dn}. 
These are also issues to be investigated in the future.

\acknowledgments
HK would like to show his appreciation to Sung-Jay Lee, Qing-Guo Huang,
Pravabati Chingangbam and Eoin \'O Colg\'ain for their advises. 
HK thanks to H. Emoto for his comments. 
HK and MS thank KIAS and its members, particularly 
Ki-Myeong Lee and Piljin Yi, for their hospitality.
The work of MN is supported in part by Grant-in-Aid for Scientific
Research No.~20740141 of Monbukagaku-sho (MEXT).
The work of MS is supported in part by JSPS Grant-in-Aid for Scientific 
Research (A) No.~21244033, and by JSPS 
Grant-in-Aid for Creative Scientific Research No.~19GS0219,
and by MEXT Grant-in-Aid for the global COE program at Kyoto University,
"The Next Generation of Physics, Spun from Universality and Emergence".


\appendix

\section{Notation}
\label{sec:notation}

\subsection{Definitions and Properties of Tensors} 
\label{definitions}
Here we explain our notation.
The Einstein tensor and the energy momentum tensor are defined as
\begin{align}
{\mathcal G}_{MN} &:= {\mathcal R}_{MN} - \frac{1}{2} g_{MN} {\mathcal R} ~~ ,&
{\mathcal T}_{MN} &:=
 - \frac{2}{\sqrt{-g}} \frac{\delta S_{_{\rm YMT}}}{\delta g^{MN}} ~~ .
\end{align}
In terms of these tensors the Einstein equation is
\begin{align}
{\mathcal G}_{MN} &=  8 \pi G {\mathcal T}_{MN} ~~ . 
\end{align}
The Einstein tensor is obtained by the differentiation of the 
Einstein-Hilbert action $S_{_{\rm EH}} $ with respect to the metric $g^{MN}$. 
The corresponding Levi-Civita connection $\Gamma^M_{NP}$ is defined as
\begin{align}
\Gamma^M_{NP} &:= \frac{1}{2} g^{MQ} \left( \partial_N g_{QP}
 + \partial_P g_{QN}  -  \partial_Q g_{NP}   \right) ~~ .
\end{align}
The Riemannian curvature ${\mathcal R}^M_{NPQ}$ is defined as
\begin{align}
{\mathcal R}^M_{NPQ} &:= \partial_P \Gamma^M_{NQ} - \partial_Q \Gamma^M_{NP} 
+ \Gamma^M_{PA} \Gamma^A_{NQ} - \Gamma^M_{QA} \Gamma^A_{NP} ~~. 
\end{align} 
The Ricci  tensor ${\mathcal R}_{MN}$  and scalar curvature ${\mathcal R}$ are 
\begin{align}
{\mathcal R}_{MN} &:= {\mathcal R}^Q_{MQN} ~~ , & 
{\mathcal R} &:= g^{MN} {\mathcal R}_{MN}  ~~. 
\end{align}

\subsection{Differential Forms}
\label{differentialform}

The tangent vector space of a point is spanned by $\partial_M$.
 The basis $dx^M$ of the cotangent space is the dual vector,
$dx^M( \partial_N ) = \delta^M_N$. For vector space $V$ the Grassmann
algebra $\Lambda^* (V)$ is defined as $T(V)/I$ where $T(V)$ is the tensor 
algebra $T(V) := \oplus_{p=0}^{\infty} V^{\otimes p}$ and $I$ is the 
two-sided ideal generated by $v \otimes v , v \in V$. 
We can define a linear operation which is called the Hodge dual.
Let us fix $p$ and $q:= D-p$. The Hodge dual operator $*$ is defined as
\begin{align}
* dX^{M_1 \cdots M_p} &:= \frac{1}{q! \sqrt{-g}}
 \epsilon^{M_1 \cdots M_p}{}_{N_1 \cdots N_q}dX^{N_1 \cdots N_q}  ~~.
\end{align}
By using the Hodge dual operation the metric on the differential
Suppose that $\omega$ is a $p$-form, 
\begin{align}
\omega &:= \frac{1}{p!} \omega_{M_1 \cdots M_p} dX^{M_1 \cdots M_p} ~~.
\end{align}
The inner product is given by $( \omega , \omega ):= \omega \wedge * \omega$.
 Let us show the metric in terms of the component, 
\begin{align}
\omega \wedge * \omega 
&= \frac{1}{(p!)^2} \omega_{M_1 \cdots M_p} 
\omega_{K_1 \cdots K_p} dX^{M_1 \cdots M_p} \wedge 
\frac{1}{q! \sqrt{-g}} 
\epsilon^{K_1 \cdots K_p}{}_{N_1 \cdots N_q}dX^{N_1 \cdots N_q}  \cr
&=  \frac{1}{(p!)^2 q! \sqrt{-g}}  \omega_{M_1 \cdots M_p} 
\omega_{K_1 \cdots K_p} \epsilon^{K_1 \cdots K_p}{}_{N_1 \cdots N_q} 
dX^{M_1 \cdots M_p N_1 \cdots N_q}  \cr
&=   - \frac{1}{g (p!)^2 q! }  \omega_{M_1 \cdots M_p} 
\omega_{K_1 \cdots K_p} \epsilon^{K_1 \cdots K_p}{}_{N_1 \cdots N_q} 
\epsilon^{M_1 \cdots M_p N_1 \cdots N_q}  dv \cr
&= -  \frac{1}{p!}  \omega_{M_1 \cdots M_p} 
\omega_{K_1 \cdots K_p} \Delta^{K_1 \cdots K_p , M_1 \cdots M_p} ~~ , 
\end{align}
where the metric $\Delta^{K_1 \cdots K_p , M_1 \cdots M_p}$ is
 defined as follows:
\begin{align}
\Delta^{K_1 \cdots K_p , M_1 \cdots M_p}
 &:= \frac{1}{p!} \sum_{\sigma \in {\mathfrak S}_p} 
{\rm sign} ( \sigma )  \prod_{i=1}^p g^{K_i M_{\sigma(i)} } ~~ . 
\end{align}
Here ${\mathfrak S}_p$ is the $p$-th symmetric group consisting of all 
permutations of $p$ characters. 
Finally we obtain
\begin{align}
\omega \wedge * \omega &= - \frac{1}{p!} \omega_{M_1 \cdots M_p} 
\omega^{M_1 \cdots M_p}  ~~.
\end{align}
The minus sign is from the fact that the signature of the metric 
$g_{MN}$  is Lorentzian. 

\subsection{Clifford algebra}
\label{clifford}

We will use the Clifford algebra with respect to the six-dimensional 
Euclidean metric in order to represent the algebra so(6).  
Indices  $a,b=1,2,\cdots ,6$ refer to the inner space.
The Clifford algebra is generated by $\gamma_a$ which satisfy
\begin{align} 
\{ \gamma_a  , \gamma_b \} &= 2 \delta_{ab} ~~,& 
\gamma_{ab} &:= \frac{1}{2} [  \gamma_a , \gamma_b ] ~~.
\end{align}
These generators are represented as $8 \times 8$ matrices. 
$\gamma_{ab}$ satisfy the commutation relation of the Lie algebra so(6). 
Anticommutation relation of $\gamma_{ab}$ is 
\begin{align}
\{ \gamma_{ab} , \gamma_{cd} \} &=  2\gamma_{abcd} - 4 \delta^{ab}_{[cd]} ~~. 
\end{align}
Here $\gamma_{abcd}$ is an antisymmetric product of four generators defined as
\begin{align}
\gamma_{a_1a_2\cdots a_p} &:= \frac{1}{p!} 
\sum_{\sigma \in {\mathfrak S}_p} {\rm sgn} \sigma 
\gamma_{a_{\sigma(1)} \cdots a_{\sigma(p)}}  ~~. 
\label{gammadef}
\end{align}
The chirality operator $\gamma_7$ is defined as
\begin{align}
\gamma_7 &= - {\bm{i}} \gamma_1\gamma_2 \gamma_3 \gamma_4 \gamma_5 \gamma_6 ~~,& 
\gamma_7^2 &=1 , & \gamma_7^{\dag} &= \gamma_7 ~~. 
\end{align}
By using this matrix, $\gamma_{abcd}$ is written as a sum of
 products of $\gamma_{7}$ and $\gamma_{ab}$, 
\begin{align}
\gamma_{abcd} &= -\frac{\bm{i}}{4!}  \epsilon_{abcdef} \gamma_7 \gamma_{ef}  ~~. 
\end{align}

\subsection{Notation for Gauge Fields}
\label{gaugefield}

The degree of freedom of a gauge boson is represented by the
 Lie algebra-valued one-form $A$,  
\begin{align}
A &:= \frac{1}{2} A_M^{ab} \gamma_{ab} dX^M , & 
F &= dA + {\bm{q}} A \wedge A . 
\end{align}
where $F$ is the corresponding gauge field strength two-form and 
${\bm{q}}$ is the  gauge coupling constant. 
Let us rewrite the action in terms of the components, 
\begin{align}
\frac{1}{16}{\rm Tr} \left( - F \wedge * F  \right)
 &= \frac{1}{32} {\rm Tr} \left(  F_{MN} F^{MN}  \right) dv\cr
&= \frac{1}{4 \cdot 32} F_{MN}^{ab} F^{cd,MN}
 {\rm Tr} \gamma_{ab} \gamma_{cd} dv \cr
&= \frac{1}{4 \cdot 32} F_{MN}^{ab} F^{cd,MN} 
dv 8 \left( \delta_{bc} \delta_{ad} - \delta_{bd}\delta_{ac} \right)  \cr
&= - \frac{1}{4 \cdot 2} F_{MN}^{ab} F^{ab,MN} dv  ~~ . 
\end{align}
For notational simplicity, we introduce the composite four form operator $H$,
\begin{align}
H &:= F \wedge F = \frac{1}{4!} H_{MNPQ} dX^{MNPQ}\,.
\end{align}
The energy momentum tensor is
\begin{align}
{\mathcal T}_{MN} &= \frac{1}{2} F_{MP}^{ab} F^{ab}_N{}^P 
+   \frac{\alpha^2}{8 \cdot 3!} {\rm Tr} H_{MPQS} H_N{}^{PQS} 
 - \frac{1}{2} g_{MN} \chi \cr
&=  \frac{1}{8} {\rm Tr}  \left( - F_{MP} F_N{}^P 
 + \frac{\alpha^2}{3!}   H_{MPQS} H_N{}^{PQS} \right)
 - \frac{1}{2} g_{MN} \chi ~~,
\end{align}
where 
\begin{align}
\chi &:= {\rm Tr} \left(  - \frac{1}{16}  F_{MN} F^{MN}
 +  \frac{\alpha^2}{4!} H_{MNPQ} H^{MNPQ} + V_0 \right)~~. 
\end{align}

\section{$\bm{\phi_2=}$constant solutions}
\label{sec:cals}

In this Appendix, we derive an inequality which gives 
the condition for 
Eq.~(\ref{eqn:solv2}) to have two real solutions. 

Because $d^2f(Z)/dZ^2=12Z^2+10\nu_1>0$ for arbitrary $Z$,  
the polynomial $f(Z)$  has a unique minimum. 
This means that the number of real solutions of $f(Z)=0$,
Eq.~(\ref{eqn:solv2}), is at most 2.
Let the value of $Z$ at the minimum be $Z_0$.
Then $f(Z_0)$ must be non-positive for a real solution to exist:
\begin{eqnarray}
f(Z_0)=Z_0^4+5\nu_1Z_0^2-32\nu_1Z_0+\frac{8}{5}\nu_1\nu_2\leq0\,.
\label{fZ0condition}
\end{eqnarray}
Also $Z_0$ must be the unique real solution of the equation,
\begin{align}
\frac{d f(Z)}{dZ} =4 Z^3 +  10 \nu_1 Z - 32 \nu_1 &=0 ~.
\label{eqn:derivf}
\end{align}
Because $- 32 \nu_1$ is negative, $Z_0$ must be positive.
In fact, by using Cardano's formula, we obtain
\begin{align}
Z_0 &=
 \left(   {4}{ \nu_1} + {4}{ \nu_1}
\sqrt{  1  +
 \frac{ 5^3}{3^3 \cdot  2^7 }\nu_1
 }  \right)^{1/3}
 -
\left(  - {4}{ \nu_1} + {4}{ \nu_1}
\sqrt{  1  +
 \frac{ 5^3}{3^3 \cdot  2^7 }\nu_1
 }  \right)^{1/3}   ~,
\end{align}
which is manifestly positive definite.

Now using $df(Z_0)/dZ=0$, the condition (\ref{fZ0condition}) reduces to
\begin{align}
   \frac{5}{2 }  Z_0^2 -{24} Z_0   &\leq -  \frac{8\nu_2}{5}~.
\label{eqn:condex}
\end{align}
Thus when the couplings $(G,V_0, {\bm{q}}, \alpha)$ satisfy the 
condition (\ref{eqn:condex}), there are one or two real 
solutions $Z_1$ and $Z_2$, ($Z_1 \geq Z_2$). Because $Z_0$ is positive,
and we have the relation $Z_1\geq Z_0\geq Z_2$,
$Z_1$ is always positive if it exists.
When the equality in Eq.~(\ref{eqn:condex}) is satisfied, we
have $Z_1=Z_2(=Z_0)$. 


We assume that the parameters satisfy Eq.~(\ref{eqn:condex}).
Then for $\nu_2 \geq 0$ or equivalently $c \geq 0$, 
we have $f(0)\geq 0$, hence both solutions are non-negative:
$Z_1 \geq Z_2 \geq 0$.
The solutions are  given by the Ferrari's formula,
\begin{align}
 Z &=   \epsilon_1  \frac{\sqrt{u}}{2}  
+ \epsilon_2 \sqrt{D}\,;
\quad
D:=- \frac{1}{4}  \left(  10  \nu_1 + u \right)  
 +    16 \epsilon_1 \frac{\nu_1}{\sqrt{u}}~.
\end{align}
Here $\epsilon_1$ and $\epsilon_2$ are $\pm 1$, and $u$ is 
\begin{align}
u &=  - \frac{10 \nu_1}{3}
+  \left(- \frac{J}{2} - \sqrt{ \frac{J^2}{4}  - \frac{H^3}{27} } \right)^{1/3}
+ \left( - \frac{J}{2} + \sqrt{ \frac{J^2}{4}  - \frac{H^3}{27} } \right)^{1/3}~,
\end{align}
where
\begin{align}
H &= \frac{25 \nu_1^2}{3} + \frac{32\nu_2\nu_1}{5} ~,&
J &=  \frac{64 \nu_1^2 \nu_2}{3} - 2^{10} \nu_1^2 -  \frac{250 \nu_1^3}{3^3}  ~.
\end{align}
If $\nu_1$ and $\nu_2$ satisfy Eq.~(\ref{eqn:condex}), 
${J^2}/{4}  - {H^3}/{27} >0 $.
This means that $u$ is positive. 
The equation must have only one or two real solutions.
This implies $\epsilon_1 = 1$ because $D<0$ if $\epsilon_1=-1$.
Thus the two real solutions are
\begin{align}
 Z_1 &=   \frac{\sqrt{u}}{2}  
+ \sqrt{ - \frac{1}{4}  \left(  10  \nu_1 + u \right)   +    
   16 \frac{\nu_1}{\sqrt{u}}  }~,&
 Z_2 &=   \frac{\sqrt{u}}{2}  
- \sqrt{ - \frac{1}{4}  \left(  10  \nu_1 + u \right)   +   
   16 \frac{\nu_1}{\sqrt{u}}  }~,
\label{Z1Z2sol}
\end{align}
where $Z_1 \geq Z_2$. 

The above expressions for the solutions $Z_1$ and $Z_2$
are quite complicated as they are. However, using the
scaling freedom of $L_0$, it is possible to simplify the
expressions. For this purpose, let us first
recapitulate Eq.~(\ref{L1def}) where the length $L_1$
representing the linear extension of the extra dimensions
was introduced,
\begin{align}
L_1^2 &= \frac{8 \pi G}{{\bf q}^2} \frac{1}{Z_1} ~.
\end{align}
Then we set $L_0 =L_1$, which implies $Z_1=a$.

Also for $Z_2$, we may also simplify the expression 
in terms of $a, b, c$ with the normalization $L_0=L_1$. 
In this case, since $Z=Z_1=a$ is a solution of $f(Z)=0$,
we have Eq.~(\ref{eq:ctoab}),
\begin{eqnarray}
f(a)=0\Leftrightarrow c=20-\frac{5a}{8}(5+63b)\,,
\label{ceq}
\end{eqnarray}
and $f(Z)$ can be divided by $(Z-a)$.
The quotient is
\begin{align}
 \frac{63ab}{8} ((Z/a)^3+(Z/a)^2)  + \left( 4 - \frac{c}{5} \right) (Z/a)
  - \frac{c}{5}  &=0 ~.
\label{eqn:eqforz2}
\end{align}
In order to use the Cardano's formula, let us change the equation into 
the normal form,
\begin{align}
(Z/a+1/3)^3 + A (Z/a+1/3) + B &=0 ~,
\end{align}
where
\begin{align}
A &= \frac{  5+42b  }{63 b}  >0 ~,&
B &= - \frac{2\bigl(48-5a(1+14b)\bigr)}{3 \cdot 63 ab} ~.
\end{align}
This equation has only one real positive solution $Z_2$.
Therefore the solution is
\begin{align}
Z_2/a &= - \frac{1}{3} - \left\{ \frac{1}{2} 
\left(  B + \sqrt{ B^2 + \frac{4A^3}{27} }  \right) \right\}^{1/3}
+  \left\{ \frac{1}{2} \left( - B + \sqrt{ B^2 + \frac{4A^3}{27} }  \right) 
  \right\}^{1/3} ~.
\end{align}

Finally let us derive the bounds on the parameters $a$ and $b$.
We assume $c\geq0$. From Eq.~(\ref{ceq}), this gives
a bound on $a$ and $b$,
\begin{eqnarray}
32-a(5+63b)>0\,.
\end{eqnarray}
In addition, since Eq.~(\ref{eqn:eqforz2}) has only one 
real positive solution $Z_2$ which is equal to or smaller than $Z_1$,
the left-hand side of it is non-negative at $Z=a$,
\begin{eqnarray}
 \frac{63ab}{4}   + \left( 4 - \frac{c}{5} \right)  
  - \frac{c}{5}  
=\frac{1}{4}\Bigl(a\left(5+126b\right)-16\Bigr)\geq 0 ~.
\label{Z1geqZ2}
\end{eqnarray}
Therefore the conditions that $Z_2 \leq Z_1$ and $c\geq0$
yield the bounds on the parameters $a$ and $b$ as
\begin{align}
\frac{ 32}{5+63b} \geq a &\geq \frac{16}{5+126 b} ~.
\label{eqn:bounda3}
\end{align}

\section{Asymptotic behavior in the case of
$\lambda_1=0$}
\label{sec:withoutav}

Here we derive the asymptotic behavior of the solution of
the system given by Eqs.~(\ref{eqn:sys3}) and (\ref{eqn:sys4}).
Equations~(\ref{eqn:r}) and (\ref{eqn:psi}) can be written as
\begin{eqnarray}
\frac{\psi'}{r}\Omega\cos(\Omega\tau+\psi)
+(\frac{r'}{r^2}+6\Omega)\Omega\sin(\Omega\tau+\psi)
&=&-3\Omega^2-\left(6\Omega^2-\frac{15}{2}\right)\cos^2(\Omega\tau+\psi),
\label{psirprime}
\\
\frac{\psi'}{r}&=&\frac{r'}{r^2}\tan^{-1}(\Omega\tau+\psi). 
\label{psiprime}
\end{eqnarray}
By eliminating the term $\psi'/r$ from these equations, we obtain
\begin{align}
\frac{r'}{r^2}   
&= -3 \Omega + {\cal F}  ~,
\label{rprimeeq}
\end{align}
where
\begin{align}
\mathcal F 
&=-  \left(\frac{9}{2} \Omega - \frac{15}{8 \Omega}  \right) \sin 
(\Omega \tau +\psi) + 3 \Omega  \cos (2\Omega \tau +2\psi)
 - \left( \frac{3}{2} \Omega - \frac{15}{8 \Omega} \right) 
   \cos (3\Omega \tau +3\psi). 
\label{mathcalF}
\end{align}
We can integrate this to obtain an expression for $r$,
\begin{align}
\frac{1}{r}   &= 
 3 \Omega \tau - \int^\tau d \tau {\cal F} ~. 
\end{align}
As for the angle $\psi$, from Eqs.~(\ref{psiprime})
and (\ref{rprimeeq}), it satisfies
\begin{align}
  \psi'   &= -r\Omega  \cos(\Omega \tau +\psi)
\left( 3 + 6 \sin(\Omega \tau +\psi)
 + \left( 6 - \frac{15}{2 \Omega^2} \right) 
 \cos^2(\Omega \tau +\psi) \right)~. 
\label{psiprimeeq}
\end{align}
As clear from this equation, $\psi$ tends to a constant for $r\to 0$,
Then $\cal F$ will be a function oscillating around zero.
This implies that the integral of ${\cal F}$ in Eq.~(\ref{rprimeeq})
cannot be large. Thus in the region where $\tau$ is large enough,
$r$ damps out in time as $1/\tau$,
\begin{align}
r &= \frac{1}{3 \Omega \tau -\int d \tau {\cal F}}
 \sim \frac{1}{3\Omega\tau}~.
\end{align}
This is consistent with our anticipation that $\psi$ tends to a
constant. 
Therefore ignoring an irrelevant integration constant,
the asymptotic behaviors of $\phi_1$ and $\phi_2$
at large $\tau$ are given by
\begin{align}
\phi_2 &\sim \frac{1}{3 \Omega \tau} \sin \Omega \tau~,& 
\phi_1 &\sim \frac{2}{3}\log\tau-{\rm Si}(\Omega \tau )~.
\end{align}


\end{document}